\documentclass[fleqn,usenatbib,useAMS]{mnras}

\usepackage{graphicx}	
\usepackage{amsmath}	
\usepackage{amssymb}	
\usepackage{multicol}        
\usepackage{bm}		
\usepackage{pdflscape}	
\usepackage{csvsimple}

\usepackage[T1]{fontenc}
\usepackage{ae,aecompl}

\usepackage{newtxtext,newtxmath}

\title[Polarisation properties of energetic pulsars]
{The Thousand-Pulsar-Array programme on MeerKAT IV: Polarisation properties of young, energetic pulsars}
\author[Serylak et al.]
{M.~Serylak$^{1}$\thanks{Email: mserylak@ska.ac.za},
S.~Johnston$^{2}$,
M.~Kramer$^{3}$,
S.~Buchner$^{1}$,
A.~Karastergiou$^{4}$,\newauthor
M.~J.~Keith$^{5}$,
A.~Parthasarathy$^{3}$,
P.~Weltevrede$^{5}$,
M.~Bailes$^{6,7}$,
E.~D.~Barr$^{3}$,\newauthor
F.~Camilo$^{1}$,
M.~Geyer$^{1}$,
B.~V.~Hugo$^{1,8}$,
A.~Jameson$^{6,7}$,
D.~J.~Reardon$^{6,7}$,\newauthor
R.~M.~Shannon$^{6,7}$,
R.~Spiewak$^{6,7}$,
W.~van Straten$^{9}$,
V.~Venkatraman~Krishnan$^{3}$
\\
$^1$South African Radio Astronomy Observatory (SARAO), 2 Fir Street, Black River Park, Observatory, Cape Town, 7925, South Africa\\
$^2$CSIRO Astronomy and Space Science, Australia Telescope National Facility, PO~Box~76, Epping NSW~1710, Australia\\
$^3$Max-Planck-Institut f\"ur Radioastronomie (MPIfR), Auf dem H\"ugel 69, D-53121 Bonn, Germany\\
$^4$Oxford Astrophysics, Denys Wilkinson Building, Keble Road, Oxford, OX1 3RH, United Kingdom\\
$^5$Jodrell Bank Centre for Astrophysics, Department of Physics and Astronomy, University of Manchester, Manchester M13 9PL, United Kingdom\\
$^{6}$Centre for Astrophysics and Supercomputing, Swinburne University of Technology, Hawthorn, VIC, 3122 Australia\\
$^{7}$Australia Research Council Centre for Excellence for Gravitational Wave Discovery (OzGrav)\\
$^{8}$Department of Physics and Electronics, Rhodes University, Artillery Road, Grahamstown, South Africa\\
$^{9}$Institute for Radio Astronomy \& Space Research, Auckland University of Technology, Private Bag 92006, Auckland 1142, New Zealand\\
}
\date{Last updated; in original form}

\pubyear{2020}

\begin{document}
\label{firstpage}
\pagerange{\pageref{firstpage}--\pageref{lastpage}}
\maketitle

\begin{abstract}
We present observations of 35 high spin-down energy radio pulsars using the MeerKAT telescope. Polarisation profiles and associated parameters are also presented. We derive the geometry for a selection of pulsars which show interpulse emission. We point out that, in several cases, these radio pulsars should also be seen in $\gamma$-rays but that improved radio timing is required to aid the high-energy detection. We discuss the relationship between the width of the radio profile and its high-energy detectability. Finally, we reflect on the correlation between the spin-down energy and the radio polarisation fraction and the implications this may have for $\gamma$-ray emission.
\end{abstract}



\section{Introduction}
In a recent paper, \citet{jskk20} examined the population of young, energetic pulsars with spin-down energy loss rates, $\dot{E}$, above 10$^{35}$~erg~s$^{-1}$. They built a model of the underlying Galactic population of these objects and determined the ratios between pulsars seen in the radio, in $\gamma$-rays or both. They showed how the numbers in each of these classes depended on both $\dot{E}$ and the geometry of the pulsar. A pulsar's geometry can be characterised by two angles, $\alpha$ the inclination angle of the magnetic axis to the rotation axis and $\beta$ the angle between the observer's line-of-sight and the magnetic axis. \citet{jskk20} demonstrated that (statistically speaking) joint radio and $\gamma$-ray pulsars have high values of $\alpha$ (i.e. were close to orthogonal) and low values of $\beta$ whereas radio-only pulsars tended to have lower values of $\alpha$. They also showed how this depends on the value of $\dot{E}$.

A radio pulsar's geometry can be determined by examination of the position angle (PA) sweep of the linear polarisation across pulse phase \citep{rc69} along with knowledge of the on-pulse width (e.g. \citealt{rwj15a}). This, in principle, can be used to determine whether the geometries of radio-only pulsars are different from the joint radio and $\gamma$-ray pulsars \citep{rwjk17}. It has long been known that young pulsars with high values of $\dot{E}$ are also highly polarised \citep{avh00,jw06,wj08a} and generally have a smooth sweep of PA, unlike pulsars at lower $\dot{E}$. There is a large body of published radio pulsar polarisation profiles, including recent compilations from \citet{wcl+99,wck+04}, \citet{hr10}, \citet{mbm+16}, \citet{jk18} and \citet{hmvd18}. Generally, these pulsars tend to be radio bright and there remains a substantial number of radio faint pulsars without polarisation measurements.

In $\gamma$-rays, the Large Area Telescope (LAT) on board the Fermi satellite \citep{fermi} has greatly increased our knowledge of the high-energy pulsars, and the current Fermi pulsar catalogue \citep{2pc} with additional updates \citep{sbc+19} lists in excess of 100 non-recycled pulsars. In spite of this, the location of the $\gamma$-ray emission remains unclear. In the outer-gap model, the emission arises high in the magnetosphere above the null charge line \citep{rom96} with other models involving curvature radiation also proposed \citep{pet19}. In the recent force-free models, $\gamma$-rays originate in an equatorial sheet beyond the light cylinder \citep{ps18,khkw19}. For a given model, it is possible to determine the pulsar geometry from the $\gamma$-ray profile as has been done for young pulsars by e.g. \citet{wrwj09} and for the millisecond pulsars by e.g. \citet{jvh+14}.

There are several aims to this paper. First, to improve the number of pulsars with polarisation properties at high $\dot{E}$. Secondly, there are a number of low-luminosity radio pulsars which have been found through deep searches of their $\gamma$-ray counterparts. Are these pulsars similar to or different from their higher luminosity counterparts? Relatively few $\gamma$-ray pulsars are detected with $\dot{E}<10^{35}$~erg~s$^{-1}$. Those with radio counterparts are expected to have high values of $\alpha$ and/or narrow pulse widths and we will test this idea. Finally, we will examine pulsars with interpulses; these orthogonal rotators should be preferentially detected in $\gamma$-rays at lower $\dot{E}$ \citep{jskk20}.

The MeerKAT telescope has the sensitivity and the capability to produce polarisation profiles for weak radio pulsars. MeerTime is the approved pulsar observing project on the MeerKAT telescope \citep{mtime}. The project is divided into four major themes, including the Thousand Pulsar Array (TPA) theme \citep{tpa20} which observes the non-recycled pulsar population. We use observations taken as part of the TPA in this paper.

\section{Source Selection}
In this paper we are interested in comparing the properties of young, non-recycled radio pulsars that are seen at $\gamma$-ray wavelengths with the properties of those that are not visible in $\gamma$-rays.

Taking only the young, non-recycled pulsars, there are 81 pulsars which emit at both radio and $\gamma$-ray wavelengths according to the public database\footnote{\href{https://confluence.slac.stanford.edu/display/GLAMCOG/Public+List+of+LAT-Detected+Gamma-Ray+Pulsars/}{https://confluence.slac.stanford.edu/display/GLAMCOG/Public+List+of+LAT-Detected+Gamma-Ray+Pulsars/}}. Of these, 49 have polarisation profiles published in the compilation of \citet{jk18}, 13 are too far north (above +30\degr\ declination) to be part of the TPA programme and the Crab pulsar was not observed. The TPA has observed the remaining 18 pulsars (classified as G in Table~\ref{tab_results}) and these results are presented here.

As a comparison set, there are 42 young pulsars with $\dot{E}>10^{35}$~erg~s$^{-1}$ which are seen only in radio. Of these, 29 have polarisation profiles published (mostly in \citealt{jk18}) and all but one of the remainder form part of the TPA with their results presented here (classified as R in Table~\ref{tab_results}). In addition we select from the lower $\dot{E}$ category five pulsars which show interpulse emission and which have no polarisation profiles in the literature.

\section{Array Calibration}
\begin{table*}
\caption{Parameters for 35 pulsars. In the ID column, N denotes non-detection, G denotes $\gamma$-ray and radio detected, R denotes radio pulsars with $\dot{E}>10^{35}$~ergs$^{-1}$ not seen in $\gamma$-rays, I denotes interpulse emission. The $M$ and $I$ superscripts denote values for the main and the interpulse. The table is also available as online supplementary material in machine-readable format.}
\label{tab_results}
\begin{center}
    \begin{tabular}{lrrrrrrrrrrrrr}
    \hline
    \hline
JNAME & $P$ & log $\dot{E}$ & dist  & ID & $T_\mathrm{obs}$ & DM            & RM             & $W_{50}$ & $W_{10}$ & \%L & \%V & \%|V| & \%err\\
      & (s) & (erg~s$^{-1}$)& (kpc) &    & (min)            & (cm$^{-3}$pc) & (rad~m$^{-2}$) & (deg)    & (deg)    &     &     &       &      \\
\hline
J0514--4408$^{M}$ & 0.320271 & 33.4 &  1.0 & GI &  30 &   15.1  &    17.4(3) &   6.3 &  34.4 &  94.4 &  --1.3 &  1.1 & 3.1\\
J0514--4408$^{I}$ &          &      &      &    &     &         &            &  27.2 &  46.4 &   5.1 &    5.0 &  7.0 & 3.2\\
J0540--6919       & 0.050569 & 38.2 & 49.7 &  G & 120 &  147.2  & --245.8    & 120.9 & 160.0 &  28.1 &    1.7 &  6.9 & 3.5\\
J0631+0646        & 0.110979 & 35.0 &  4.6 &  G &  45 &  195.0  &   105(1)   &  66.1 &       &  54.8 &    7.6 &  9.2 & 4.0\\
J0633+1746        & 0.237099 & 34.5 &  0.2 & NG &   5\\
J0835--3707$^{M}$ & 0.541404 & 33.4 &  0.6 &  I &  10 &  112.3  &    62.8(7) &   2.1 &   6.7 &  13.2 &    5.0 &  6.3 & 3.0\\
J0835--3707$^{I}$ &          &      &      &    &     &         &            &   4.2\\
J1124--5916       & 0.135477 & 37.1 &  5.0 &  G & 120 &  329.2  &   164.0(6) &  22.5 &       & 101.2 &  --3.2 &  4.9 & 3.6\\
J1151--6108       & 0.101633 & 35.6 &  2.2 &  G &  60 &  217.8  &   183.1(6) &  21.1 &  48.0 &  63.8 &  --0.8 &  7.1 & 3.2\\
J1400--6325       & 0.031182 & 37.7 &  7.0 &  R &  20 &  563.0\\
J1437--5959       & 0.061696 & 36.1 &  8.5 &  R &  60 &  549.6  & --705(10)  &  25.3 &  65.0 &  56.6 &    9.6 &  6.2 & 3.5\\
J1732--3131       & 0.196543 & 35.2 &  0.6 & NG &  10\\
J1741--2054       & 0.413700 & 34.0 &  0.3 &  G &  40 &    4.7  &            &  21.1 &  40.0 &  15.8 &   15.3 & 11.6 & 3.8\\
J1747--2809       & 0.052153 & 37.6 &  8.1 & NR &  60\\
J1747--2958       & 0.098814 & 36.4 &  2.5 &  G &  60 &  101.5\\
J1755--0903$^{M}$ & 0.190710 & 33.6 &  0.2 &  I &  10 &   63.7  &    89.2(2) &   7.0 &  17.6 &  29.0 &    5.9 & 12.1 & 3.0\\
J1755--0903$^{I}$ &                 &      &    &     &         &    24.0    &       &  25.7 &   4.0 &    6.2 &  3.6\\
J1816--0755$^{M}$ & 0.217643 & 34.4 &  3.1 & GI &   5 &  117.7  &    28.0(6) &   7.4 &  14.4 &  12.4 &  --4.0 &  4.5 & 3.0\\
J1816--0755$^{I}$ &          &      &      &    &     &         &            &   7.9 &       &   4.8 &    4.0 &  6.0 & 3.7\\
J1833--1034       & 0.061884 & 37.5 &  4.1 &  G & 120 &  169.5  &    60(4)   &  14.1 &       &  50.8 &  --8.6 &  5.5 & 3.7\\
J1843--0702$^{M}$ & 0.191615 & 34.1 &  4.3 &  I &  30 &  228.6  &   186(1)   &   6.3 &  14.1 &  14.1 &    0.4 &  2.5 & 3.1\\
J1843--0702$^{I}$ &          &      &      &    &     &         &            &   9.3 &  18.8 &  31.2 &  --8.2 &  4.5 & 3.5\\
J1849+0409$^{M}$  & 0.761194 & 33.4 &  1.7 &  I &  20 &  63.97  &    19.5(2) &   3.5 &   7.4 &  53.4 &   23.3 & 23.4 & 3.1\\
J1849+0409$^{I}$  &          &      &      &    &     &         &            &   2.1 &  13.4 &  29.0 &  --0.8 &  2.7 & 3.1\\
J1850--0026       & 0.166634 & 35.5 &  6.7 &  R &  15 &  947.0  &   664(1)   &  36.6 &       &  48.4 & --10.9 &  8.0 & 3.0\\
J1856+0113        & 0.267440 & 35.6 &  3.3 &  G &   5 &   96.1  & --122.0(3) &   3.2 &   8.4 &  57.1 &  --1.0 &  1.1 & 3.1\\
J1856+0245        & 0.080907 & 36.7 &  6.3 &  R &  30 &  623.5  &   255(2)   &  75.9 &       &  61.7 &   26.0 & 20.5 & 3.2\\
J1857+0143        & 0.139760 & 35.7 &  4.6 &  G &  30 &  249.4  &    29.4(3) &  40.8 &       &  68.4 &    3.4 &  2.9 & 3.1\\
J1906+0746        & 0.144073 & 35.4 &  7.4 & NR &   3\\
J1907+0602        & 0.106633 & 36.4 &  2.4 & NG &  10\\
J1907+0631        & 0.323648 & 35.7 &  3.4 &  R &  30 &  429.4  &   435.5(3) &  17.6 &  42.9 &  87.0 &  --9.5 &  7.3 & 3.2\\
J1907+0918        & 0.226107 & 35.5 &  8.2 &  R &  15 &  357.8  &   688.8(2) &   2.5 &   6.3 &  61.5 &   47.4 & 47.8 & 3.1\\
J1909+0749$^{M}$  & 0.237161 & 35.7 &  8.4 & RI &  30 &  539.3  & --240.9(3) &   4.9 &  16.2 &  40.6 &  --8.4 &  5.1 & 3.6\\
J1909+0749$^{I}$  &          &      &      &    &     &         &            &   7.2 &  17.6 &  86.0 &   13.8 &  7.0 & 3.7\\
J1918+1541$^{M}$  & 0.370883 & 33.3 &  0.8 &  I &  30 &   11.5  &   --5.0(8) &   4.2 &  24.6 &  66.1 &   16.9 & 18.9 & 3.3\\
J1918+1541$^{I}$  &          &      &      &    &     &         &            &  12.0 &       &  17.0 &  --5.3 &  2.7 & 4.2\\
J1925+1720        & 0.075659 & 36.0 &  5.1 &  G &  60 &  222.3  &   445.8(6) &  11.6 &       &  69.1 &    8.0 & 12.8 & 4.4\\
J1928+1746        & 0.068730 & 36.2 &  4.3 &  G &  30 &  176.7  &   203.2(2) &  17.6 &  57.0 &  50.1 & --25.9 & 26.9 & 3.2\\
J1930+1852        & 0.136855 & 37.1 &  7.0 &  R & 150 &  307.3  &            &  50.6\\
J1932+2220        & 0.144470 & 35.9 & 10.9 &  G &  15 &  218.9  &   138.9(1) &   3.2 &  10.9 &  76.5 &    7.0 &  6.6 & 3.0\\
J1934+2352        & 0.178432 & 36.0 & 12.2 &  R &  30 &  355.8  &  --35.4(6) &   7.4 &  22.9 &  71.6 &  --3.4 &  2.0 & 3.2\\
J1938+2213        & 0.166116 & 35.6 &  3.4 &  R &  15 &   93.0  &   140.5(2) &   9.8 &  28.1 &  65.1 &    8.3 &  6.4 & 3.0\\
J2043+2740        & 0.096131 & 34.7 &  1.5 &  G &  30 &   21.04 &  --96.1(1) &   4.6 &  16.5 &  84.8 &  --6.9 &  6.8 & 3.0\\
\hline
    \end{tabular}
\end{center}
\end{table*}
MeerKAT is an interferometric array located in the Karoo, in South Africa's Northern Cape Province \citep{jon16}, and consists of 64 unblocked aperture offset Gregorian antennas. It can be used to observe pulsars, and a comprehensive description of that observing system can be found in \citet{mtime}. Here we focus on a detailed description of the configuration steps relevant to creating a tied-array beam (TAB) and polarisation calibration of the beamformed data. For an exhaustive description of calibrating interferometric arrays we refer to \citet{smi11} and \citet{hal17}.

After successful array initialisation, which includes activating antennas, selecting the frequency band and the number of channels, and activation of the beamformer and pulsar backend, a set of imaging type observations is performed before each session. These observations are performed in stages, during which all of the corrections made to the individual antenna streams are tied to a reference antenna selected by the calibration pipeline. The choice of reference antenna is based on performing the Fast Fourier Transform (FFT) of imaging cross-polarisation data over all antenna pairs (baselines) and selecting the antenna with the maximum peak-to-noise ratio in the FFT spectrum.

The first calibration step, {\it delay calibration}, is performed as follows. First, predefined complex gain values are applied in the F-engine (responsible for gathering and channelizing the data streams from all antennas). Subsequently the array is pointed at a well known, stable calibrator, either PKS~J0408--6545, PKS~J0825--5010 or PKS~J1939--6342, and noise diodes are turned on to emit continuously for the duration of the track. Upon completion of the track an automated calibration pipeline is activated and \emph{calibration products} are calculated. Specifically, these products are: antenna-based geometrical delays (\textbf{\textit{K}} solutions), per-antenna bandpass corrections (\textbf{\textit{B}} solutions), per-antenna complex gain corrections (\textbf{\textit{G}} solutions), per-antenna cross-polarisation delays due to the nanosecond offsets between digitised streams (\textbf{\textit{KCROSS}} solutions) and cross-polarisation phase (\textbf{\textit{BCROSS\_SKY}} solution). It must be noted however, that only \textbf{\textit{K}} and \textbf{\textit{KCROSS}} solutions are combined and applied to the F-engine streams at this stage. The rest of the calibration solutions are stored in the observation metadata. The observation is concluded after performing another short track repeating the previous observing sequence in order to verify the accuracy of the calibration solution.

The second calibration step is designed to be used before pulsar observations and is called {\it phase up}. Similar to {\it delay calibration}, it is also divided into two tracks during which the array is pointed at a calibrator, and noise diodes are activated with the same calibration pipeline operation sequence. Specifically, the pipeline re-derives all previously mentioned calibration products (\textbf{\textit{K}}, \textbf{\textit{B}}, \textbf{\textit{G}}, \textbf{\textit{KCROSS}}, \textbf{\textit{BCROSS\_SKY}} solution), but this time the \textbf{\textit{K}} and \textbf{\textit{KCROSS}} solutions are applied as differential (fine) corrections to the {\it delay calibration} (coarse correction) solution. The rest of the calibration products are combined in per-antenna, frequency-resolved F-engine complex-valued arrays (F-engine corrections) and applied to each of the antenna data streams. This, on top of phase correction ensures that the bandpass is also corrected, essentially making it "flat" across the whole usable band. The resulting F-engine data stream is then sent to the beamformer (B-engine) which coherently adds it and forms a single TAB data stream that is then received and further processed by the pulsar backend.

We note however, that prior to 2020 April 9, the \textbf{\textit{BCROSS\_SKY}} calibration solution was calculated and applied offline to the pulsar data for every MeerTime observation. This was done by using the intermediate \textbf{\textit{BCROSS}} calibration solution calculated by the calibration pipeline and stored in the {\it delay calibration} observation metadata. It corrects for cross-polarisation phase introduced in the telescope signal chain after the noise diodes (which are placed in the receivers assemblies). The second part of the \textbf{\textit{BCROSS\_SKY}} solution, calculated offline, corrects for the cross-polarisation phase introduced by the antenna structure (primary and secondary reflectors as well as part of receiver structure, e.g. feed horn) and performs the absolute alignment of linear polarisation. Once combined into a \textbf{\textit{BCROSS\_SKY}} solution it was converted to an appropriate Jones matrix (diagonal terms being $e^{i \rho}$ and 1 and off-diagonal terms set to 0, where $\rho$ is the cross-polarisation phase) and written to a correction file that was then applied to the pulsar data using the \texttt{pac} routine from the \textsc{psrchive}\footnote{\href{http://psrchive.sourceforge.net}{http://psrchive.sourceforge.net}} software \citep{hwm04}. From 2020 April 9 onward, the observations had the full \textbf{\textit{BCROSS\_SKY}} solution applied by the automatic calibration pipeline.

As was shown in \citet{mtime}, the MeerKAT L-band receiver characteristics pertaining to polarisation purity: ellipticity and non-orthogonality (which also includes differential ellipticity) are close to ideal. Both parameters characterising the degree of mixing between linear and circular polarisation and orientation of both receptors with respect to each other (ideally both dipoles should be oriented at 0$^{\circ}$ and 90$^{\circ}$) respectively. The antenna-based leakage terms, describing imperfections in the response of the system to a polarised signal, are negligible, which combined with the calibration method described above is sufficient to obtain polarisation calibrated data and including the absolute polarisation position angles after correction for the parallactic angle.

\section{Observations and Data Analysis}
A total of 35 pulsars were observed with MeerKAT for this project and carried out as described in \citet{tpa20}. Five pulsars were not detected. PSR~J0633+1746, also known as Geminga, has been known for many years as a high-energy pulsar \citep{hh92} but has only been seen at very low radio frequencies \citep{mm97}. PSR~J1732--3131 also has only a tentative detection at very low radio frequencies \citep{mkn+17}. PSR~J1747--2809 is a weak pulsar with a high DM \citep{crgl09}. In spite of 60 minutes observing the pulsar is not detected, most likely due to high levels of scattering. PSR~J1907+0602 is a very faint radio pulsar with only a weak detection after 2 hours observing with the Arecibo telescope \citep{abdo1907}. Our short observation failed to detect the pulsar. PSR~J1906+0746 is in a binary system and its current precessional phase makes its flux density very low. Comprehensive polarisation studies are presented in \citet{dkl+19}.

In Table~\ref{tab_results}, the first four columns show the pulsars observed, their spin-period, spin-down energy and distance taken from the pulsar catalogue\footnote{\href{http://www.atnf.csiro.au/research/pulsar/psrcat}{http://www.atnf.csiro.au/research/pulsar/psrcat}} \citep{mht+05}. Column 5 gives an ID where N denotes non-detection, G denotes $\gamma$-ray and radio detected, R denotes radio pulsars with $\dot{E}>10^{35}$~ergs$^{-1}$ not seen in $\gamma$-rays, I denotes interpulse emission as per the source selection described in Section~2. Column 6 gives the total observing time in minutes.

For each pulsar we detected, we compute the dispersion measure (DM) and the rotation measure (RM) in the following way. First we create a single noise-free template of the pulsar's profile using the \textsc{ psrchive} \citep{hwm04} routine \texttt{paas}. We then sum the data in frequency using the nominal DM and compute a time-of-arrival (ToA) for each 8-second time interval using the routine {\sc pat}. If required, we use the {\sc tempo2} software\footnote{\href{https://www.atnf.csiro.au/research/pulsar/tempo2}{https://www.atnf.csiro.au/research/pulsar/tempo2}} \citep{hem06} to update the ephemeris to correct for any phase drift during the observation. We then revert to the original data set, sum the data in time, and reduce the number of frequency channels to 32. For each frequency channel we compute a ToA using the noise-free template and the ToAs are then are fitted to a DM model. The DM as measured is therefore the value which minimises the residuals for this particular template. Once the correct DM is established we then use the same 32 channel data and the routine \texttt{rmfit} to calculate the RM. This routine performs trial RMs and finds the value at which the linear polarisation is maximised across the profile as a whole. Results are listed in Table~\ref{tab_results} with the errors on the last digit of RM given in brackets. For the DM, we list to one decimal place even though the formal statistical error is usually an order of magnitude lower. We note there is a relationship between the RM and DM values \citep{okj20}.

For each pulsar we measure the profile width at 50\% of the peak amplitude ($W_{50}$) and at 10\% of the peak amplitude ($W_{10}$) where the signal-to-noise permits. Errors on $W_{50}$ and $W_{10}$ are 0.2\degr. The percentage of linear polarisation ($L/I$), of circular polarisation ($V/I$) , and of the absolute value of the circular polarisation ($|V|$) is measured for the integrated profile. To debias the linear polarisation we use equation~11 in \cite{ew01}. We follow the prescription given in \cite{kj04} to compute $|V|$ which ensures that the off-pulse baseline retains a mean of zero. For pulsars with interpulses we compute the parameters separately for both main and interpulse. The final columns of Table~\ref{tab_results} list $W_{50}$, $W_{10}$ and the polarisation parameters. The error on the polarisation fraction is the statistical error added quadrature with a conservative estimate of 3\% for systematic effects.

\section{Pulsar Profiles}
\begin{figure*}
\begin{center}
\begin{tabular}{ccc}
\includegraphics[width=5cm,angle=0]{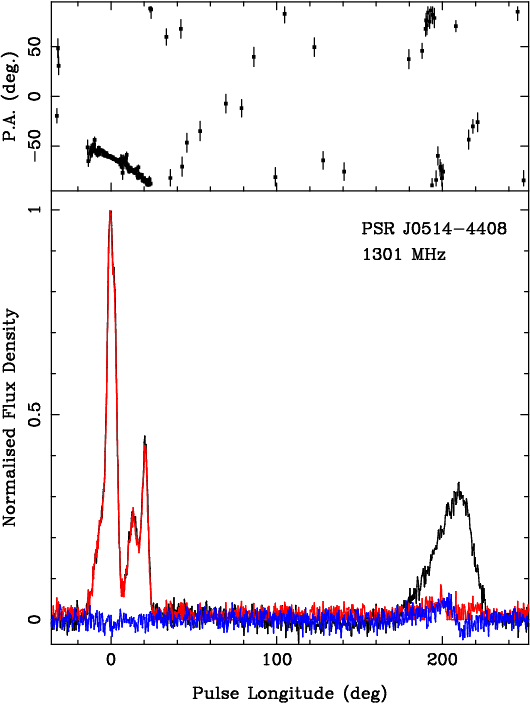} &
\includegraphics[width=5cm,angle=0]{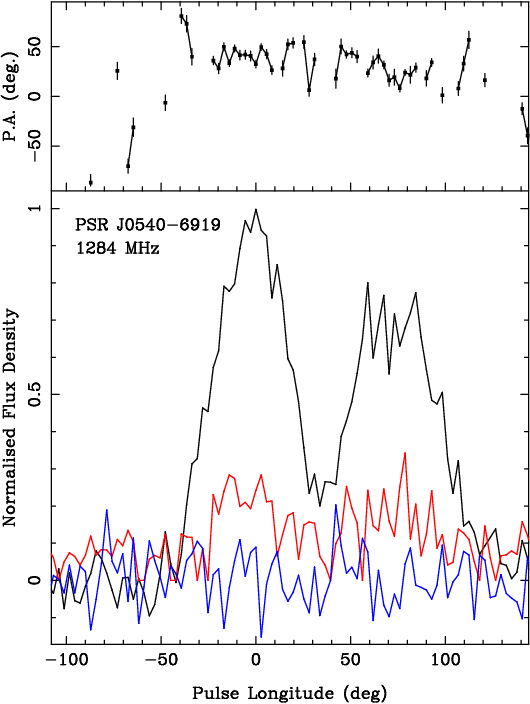} &
\includegraphics[width=5cm,angle=0]{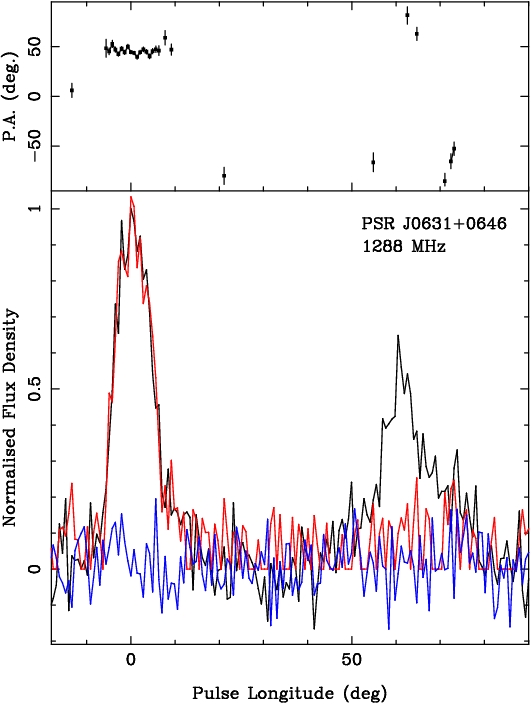} \\
\includegraphics[width=5cm,angle=0]{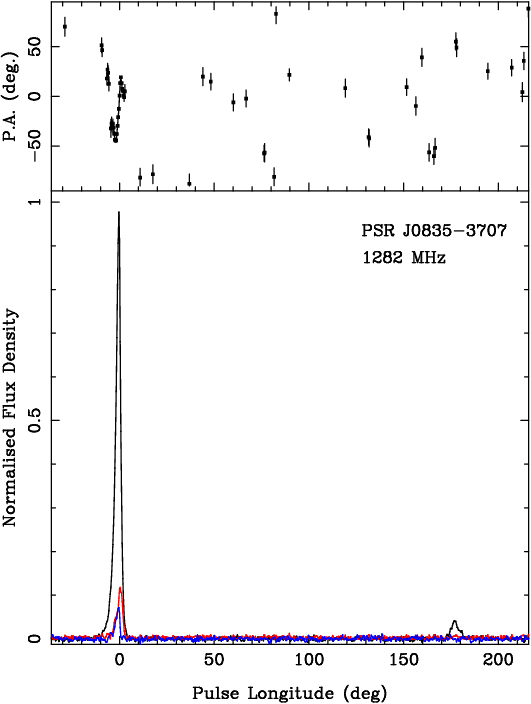} &
\includegraphics[width=5cm,angle=0]{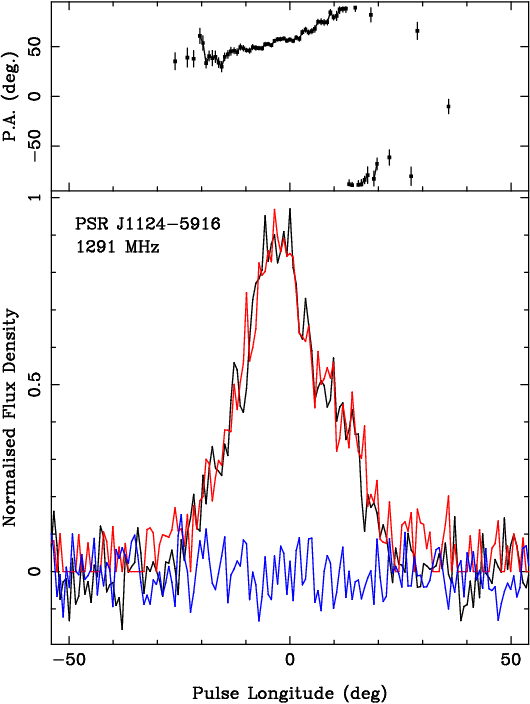} &
\includegraphics[width=5cm,angle=0]{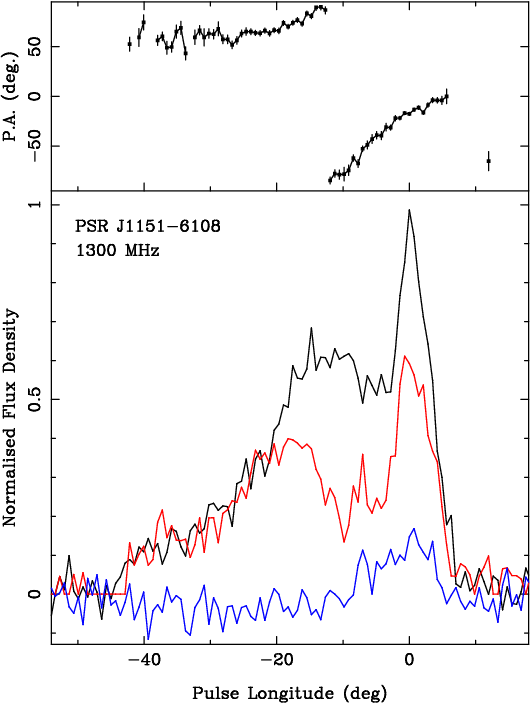} \\
\includegraphics[width=5cm,angle=0]{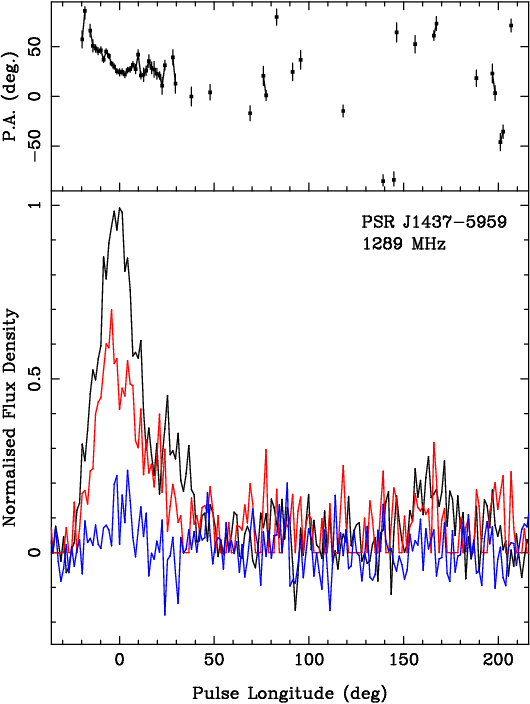} &
\includegraphics[width=5cm,angle=0]{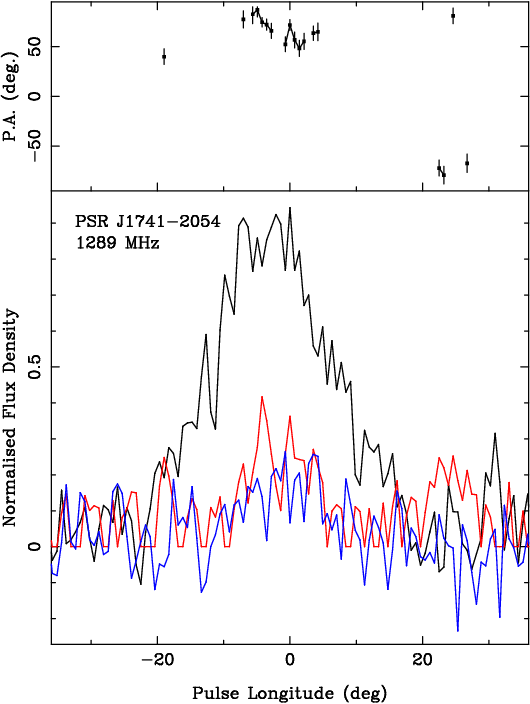} &
\includegraphics[width=5cm,angle=0]{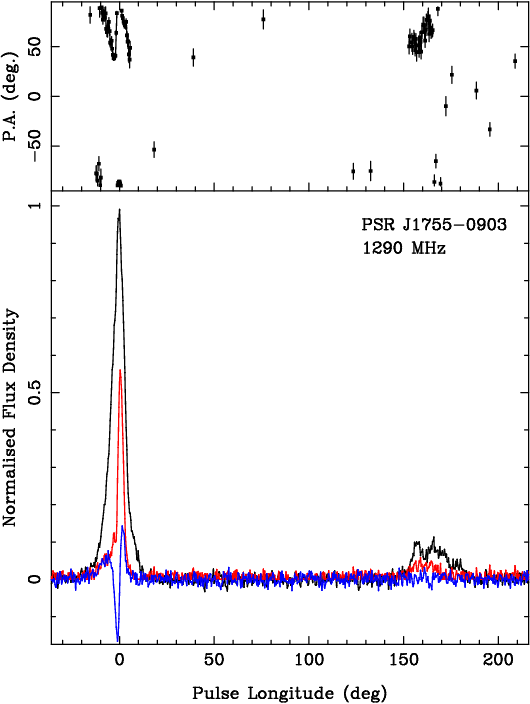} \\
\end{tabular}
\end{center}
\caption{Polarisation profiles for PSRs~J0514--4408, J0540--6919, J0631+0646, J0835--3707, J1124--5916, J1151--6108, J1437--5959, J1741--2054 and J1755--0903. In the lower panels, the black line denotes Stokes I, the red trace shows the linear polarisation and the blue trace the circular polarisation. Left-hand circular polarisation is defined to be positive. The top panel shows the position angle of the linear polarisation, corrected to infinite frequency using the RM listed in Table~\ref{tab_results}. Position angles are only plotted when the linear polarisation exceeds 3 sigma. The zero point of pulse longitude is set to the peak of the total intensity profile.}
\label{fig1}
\end{figure*}

\begin{figure*}
\begin{center}
\begin{tabular}{ccc}
\includegraphics[width=5cm,angle=0]{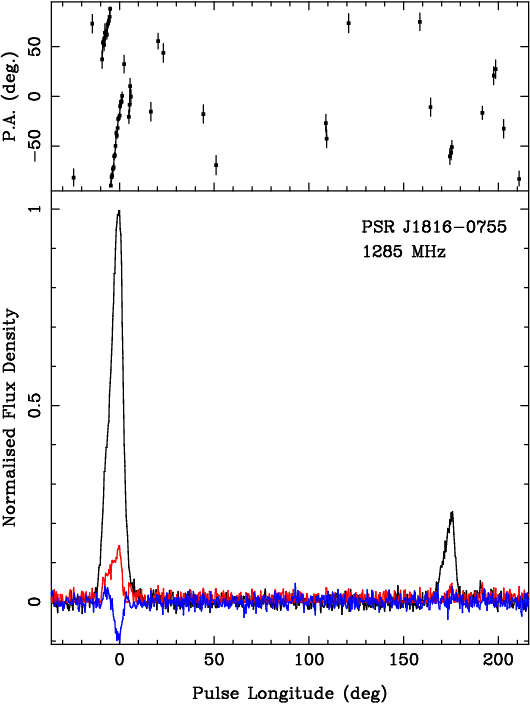} &
\includegraphics[width=5cm,angle=0]{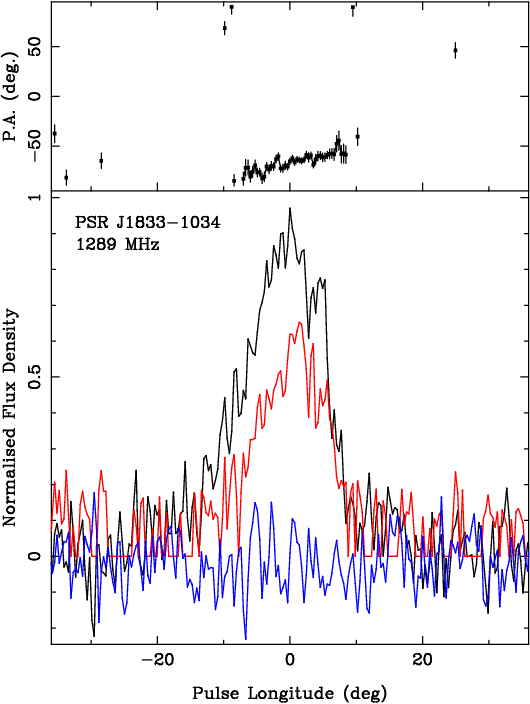} &
\includegraphics[width=5cm,angle=0]{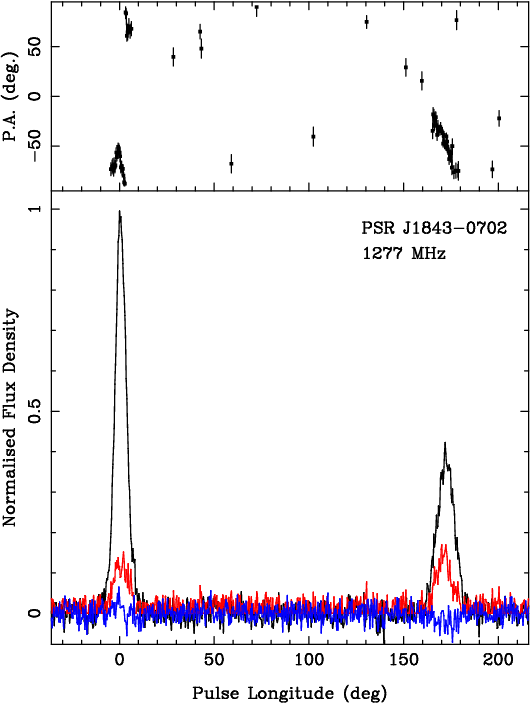} \\
\includegraphics[width=5cm,angle=0]{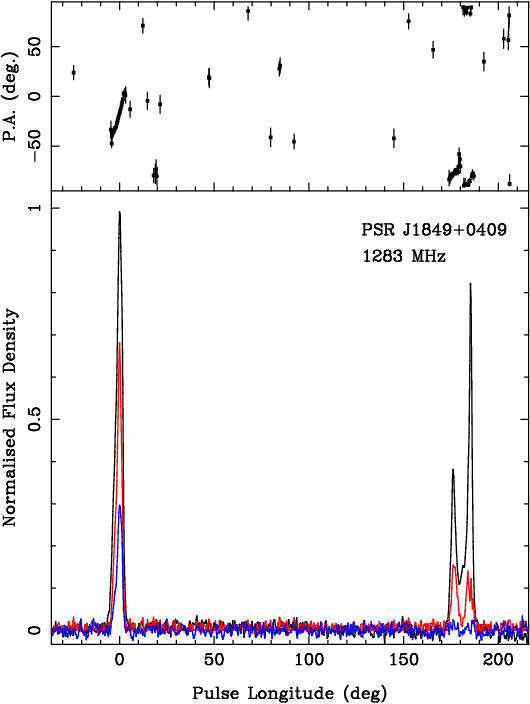} &
\includegraphics[width=5cm,angle=0]{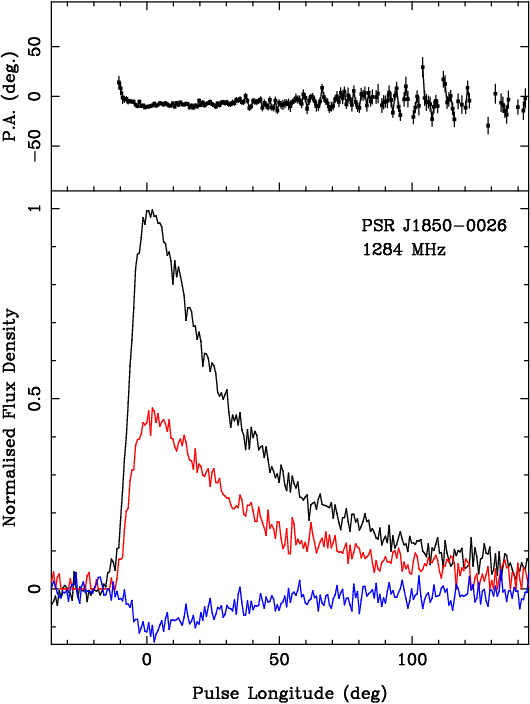} &
\includegraphics[width=5cm,angle=0]{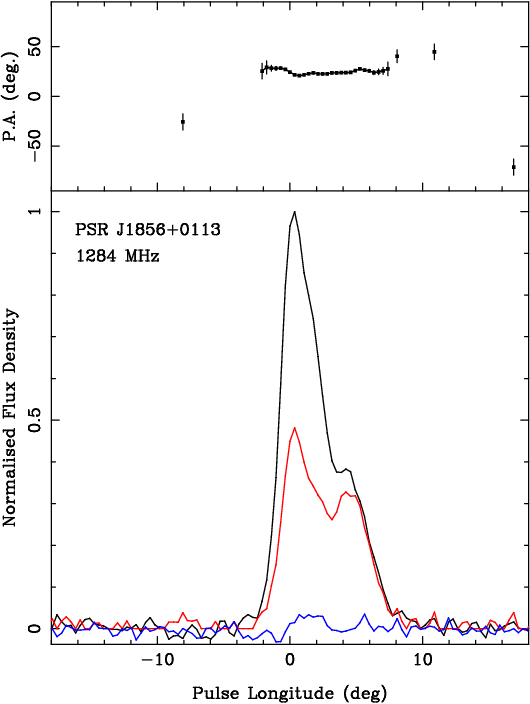} \\
\includegraphics[width=5cm,angle=0]{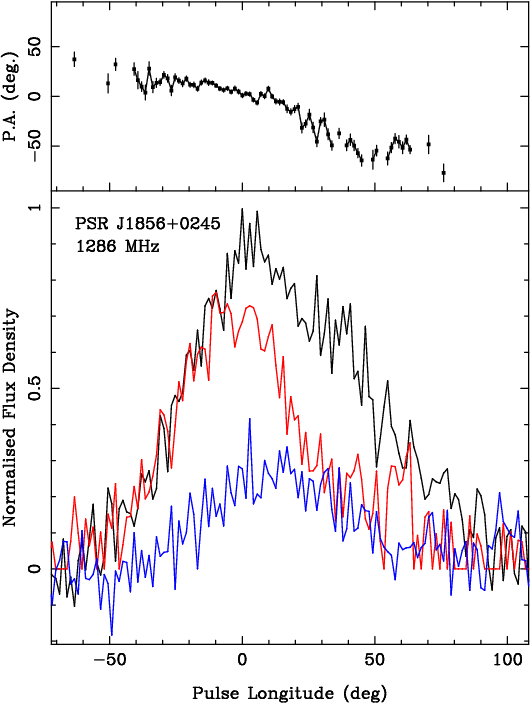} &
\includegraphics[width=5cm,angle=0]{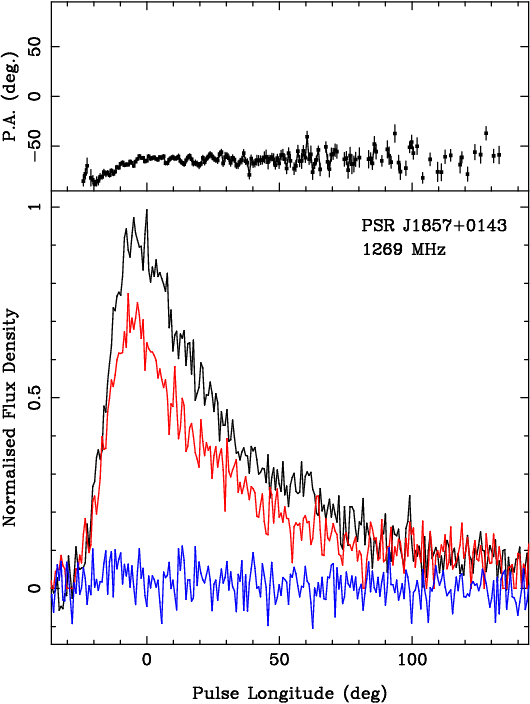} &
\includegraphics[width=5cm,angle=0]{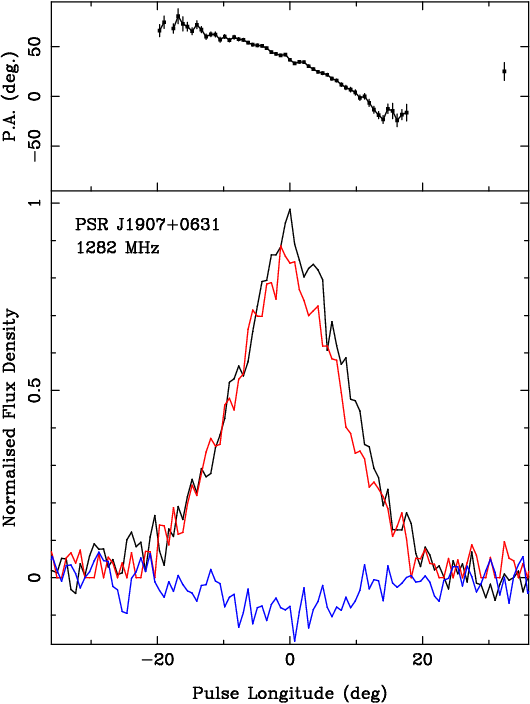} \\
\end{tabular}
\end{center}
\caption{Polarisation profiles for PSRs~J1816--0755, J1833--1034, J1843--0702, J1849+0409, J1850--0026, J1856+0113, J1856+0245, J1857+0143 and J1907+0631. In the lower panels, the black line denotes Stokes I, the red trace shows the linear polarisation and the blue trace the circular polarisation. Left-hand circular polarisation is defined to be positive. The top panel shows the position angle of the linear polarisation, corrected to infinite frequency using the RM listed in Table~\ref{tab_results}. Position angles are only plotted when the linear polarisation exceeds 3 sigma. The zero point of pulse longitude is set to the peak of the total intensity profile.}
\label{fig2}
\end{figure*}

\begin{figure*}
\begin{center}
\begin{tabular}{ccc}
\includegraphics[width=5cm,angle=0]{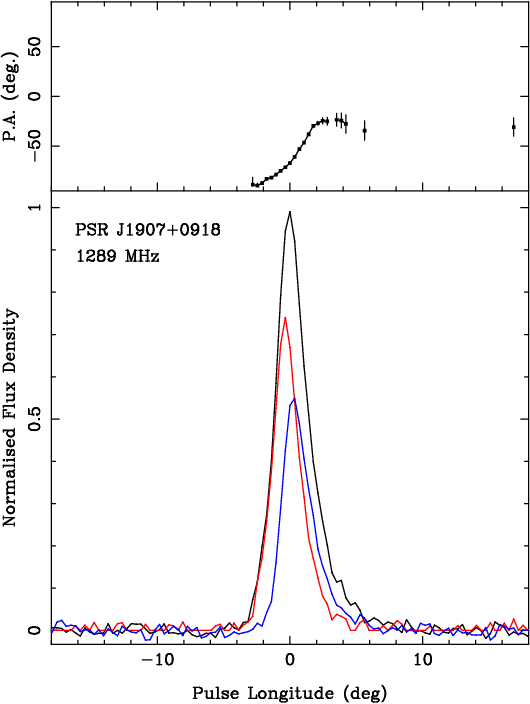} &
\includegraphics[width=5cm,angle=0]{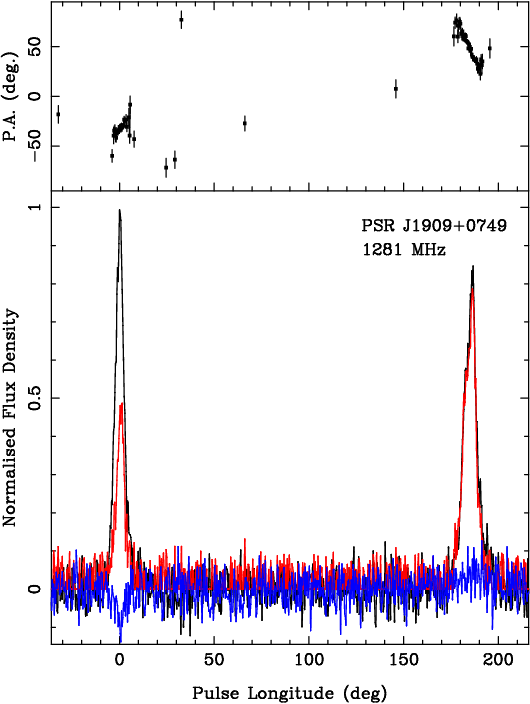} &
\includegraphics[width=5cm,angle=0]{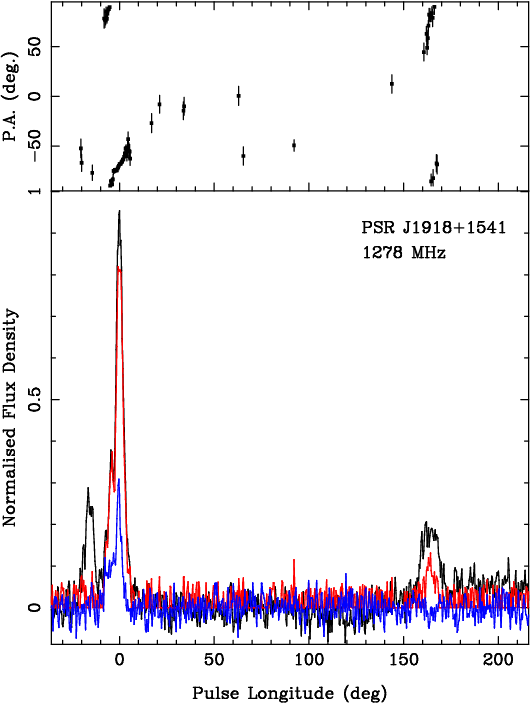} \\
\includegraphics[width=5cm,angle=0]{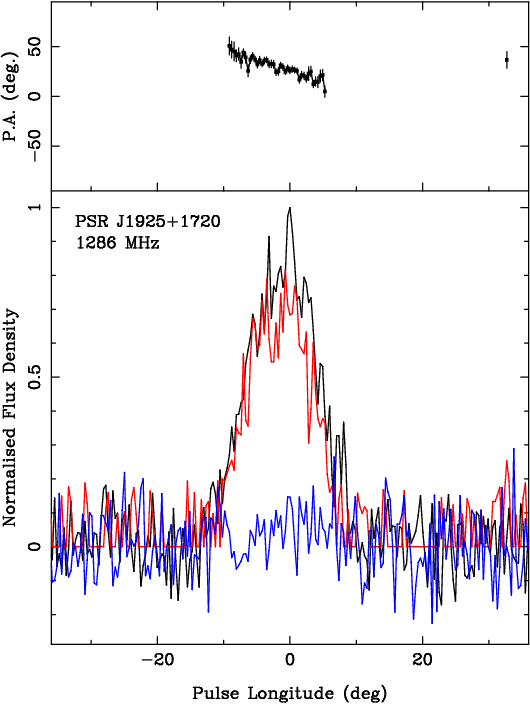} &
\includegraphics[width=5cm,angle=0]{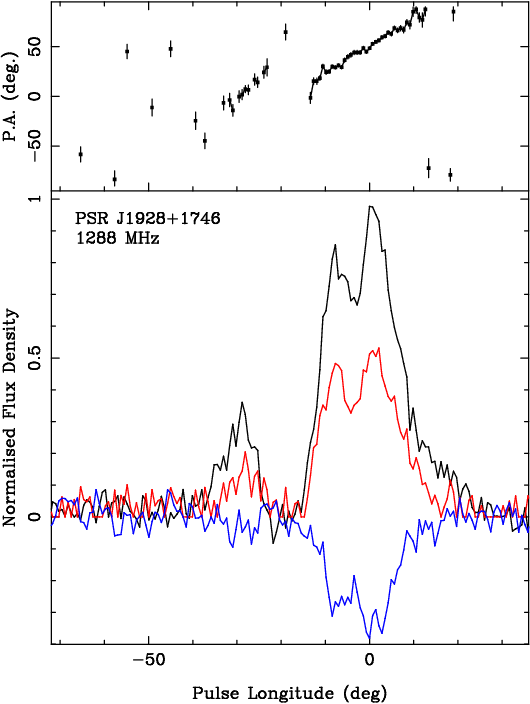} &
\includegraphics[width=5cm,angle=0]{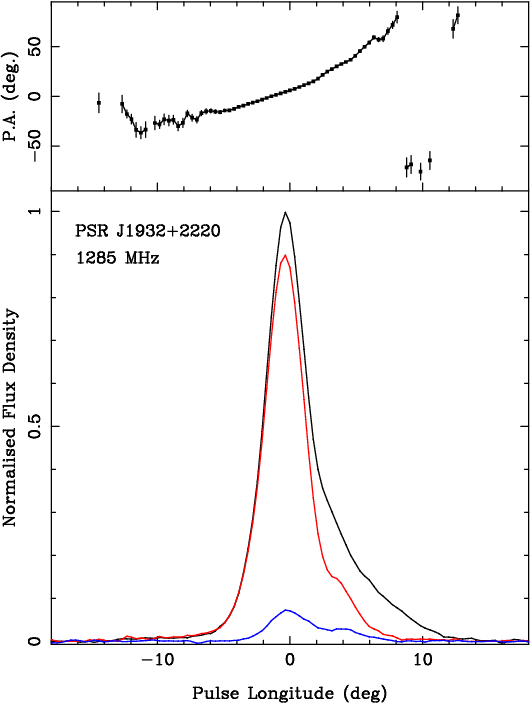} \\
\includegraphics[width=5cm,angle=0]{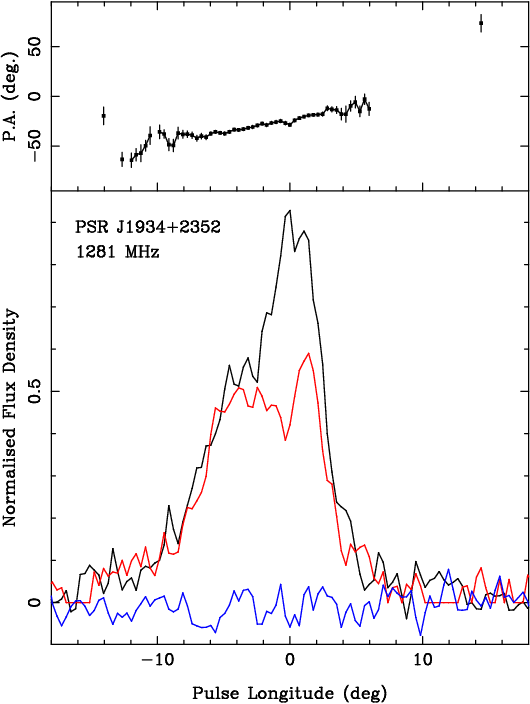} &
\includegraphics[width=5cm,angle=0]{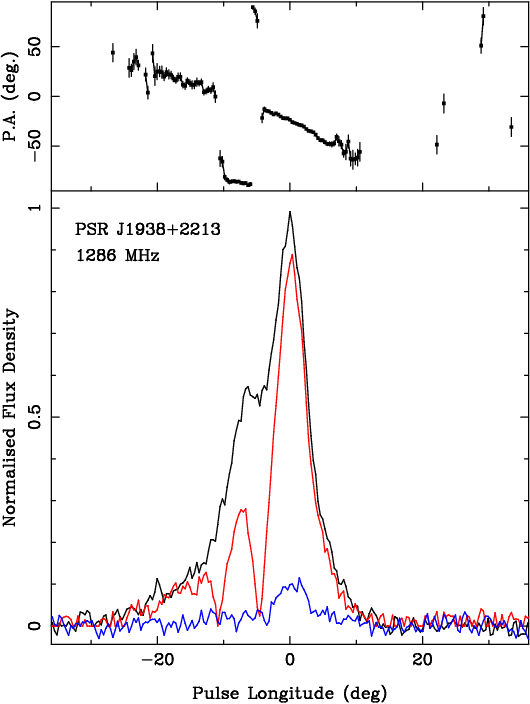} &
\includegraphics[width=5cm,angle=0]{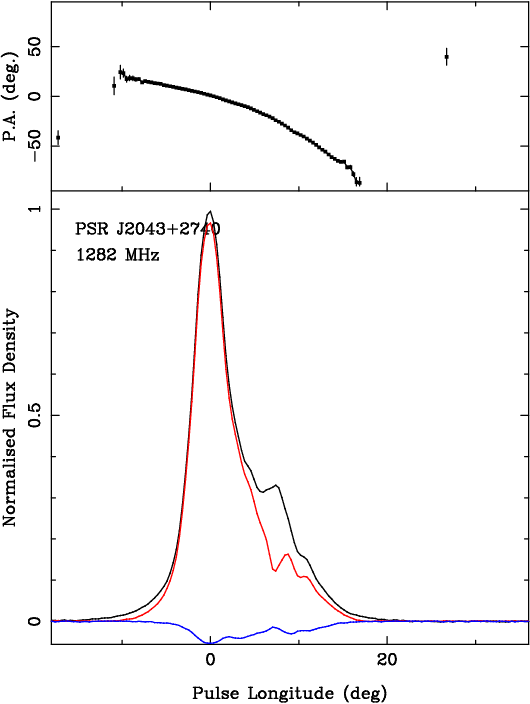} \\
\end{tabular}
\end{center}
\caption{Polarisation profiles for PSRs~J1907+0918, J1909+0749, J1918+1541, J1925+1720, J1928+1746, J1932+2220, J1934+2352, J1938+2213 and J2043+2740. In the lower panels, the black line denotes Stokes I, the red trace shows the linear polarisation and the blue trace the circular polarisation. Left-hand circular polarisation is defined to be positive. The top panel shows the position angle of the linear polarisation, corrected to infinite frequency using the RM listed in Table~\ref{tab_results}. Position angles are only plotted when the linear polarisation exceeds 3 sigma. The zero point of pulse longitude is set to the peak of the total intensity profile.}
\label{fig3}
\end{figure*}

The pulsar profiles are shown in Figures~\ref{fig1} through \ref{fig3}.
Position angles are defined as increasing counter-clockwise on the sky (see \citealt{ew01}) and are corrected to infinite frequency using the RM given in Table~\ref{tab_results}. Circular polarisation in pulsar astronomy uses the IEEE convention for left-hand and right-hand and so in the profiles shown here, left-hand circular polarisation is positive \citep{vmjr10}. A brief description for each of the detected pulsars follows below.


\noindent
{\bf J0514--4408} (Figure~\ref{fig1}): The polarisation profile for this pulsar was published by \citet{brs+19} and we get very similar results. The main component is virtually 100\% linearly polarised and consists of at least three narrow components. In contrast the interpulse has hardly any linear polarisation and has a single, wide component. The slope of the PA swing is negative against the main pulse and positive against the interpulse.

\noindent
{\bf J0540--6919} (Figure~\ref{fig1}): This pulsar is located in a supernova remnant in the Large Magellanic Cloud and is in many ways a twin of the Crab pulsar. The radio profile consists of two broad components and the overall pulse width is extremely wide. The polarisation is unusually low for such a high $\dot{E}$ pulsar and the swing of PA is remarkably flat across the entire pulse width \citep{gs20}.
 
\noindent
{\bf J0631+0646} (Figure~\ref{fig1}): The pulsar was first discovered in $\gamma$-rays by \citet{cwp+17} and subsequently as a weak radio pulsar by \citet{wcp+18}. The unusual profile consists of two widely separated components. The linear polarisation is very high in the leading component but appears to be almost zero in the trailing component. The PA swing is flat across the leading component. It is possible that the highly polarized component arises from much higher in the magnetosphere than the trailing component.

\noindent
{\bf J0835--3707} (Figure~\ref{fig1}): This pulsar has an extreme ratio between the amplitudes of the main and interpulses, and was not recognised as an interpulse pulsar in the discovery paper \citep{mlc+01}. The main pulse is very narrow, with a low level of linear and circular polarisation. The PA swing is complex and perhaps contains orthogonal mode jumps. The interpulse has a peak flux density a factor of 20 lower than the main pulse and seems to have virtually no polarisation.

\noindent
{\bf J1124--5916} (Figure~\ref{fig1}): This pulsar is located in the SNR~G292.0+1.8 and has a flux density of only 80~$\mu$Jy \citep{cmg+02}. The profile appears to be a simple Gaussian with a moderate width. It is virtually 100\% linearly polarised with almost no circular polarisation. The swing of PA is steep.

\noindent
{\bf J1151--6108} (Figure~\ref{fig1}): This pulsar was discovered by \citet{ncb+15} but has no published polarisation data. The profile is broad with a shallow leading edge and a steep trailing edge. The linear polarisation is high, the PA swing is smooth with an inflexion point near the centre of the profile. The profile follows the young pulsar type as noted by \citet{jw06} with the trailing component being the largest amplitude and showing circular polarisation.

\noindent
{\bf J1400--6325} This pulsar was first discovered in X-rays; the subsequent radio detection presented in \citet{rmg+10} shows a very broad profile at 1.4~GHz. Our observations (not shown) have a very similar profile with a width which covers more than half of the pulse period. This does not seem due to interstellar scattering as also pointed out by \citet{rmg+10}. The low signal-to-noise ratio of our profile means we were unable to determine an RM.

\noindent
{\bf J1437--5959} (Figure~\ref{fig1}): The pulsar is located in the supernova remnant 315.9--0.0. Observations made by \citet{cng+09} listed 50\% linear polarisation but the polarised profile was not shown. In our observations, we confirm the presence of interpulse emission located 170\degr\ away from the main pulse. The main pulse is broad with two components, the polarisation fraction is high and we confirm the RM given in \citet{cng+09}.

\noindent
{\bf J1741--2054} (Figure~\ref{fig1}): This pulsar has a very low flux density and a small DM and is therefore one of the lowest luminosity pulsars known in the radio \citep{crr+09}. A polarisation profile has not been published. In our observations, the profile appears to be a simple Gaussian and the polarisation fraction is low.

\noindent
{\bf J1747--2958} This pulsar is located in the radio nebula G359.23--0.82 \citep{cmgl02} and has a very low flux density. Our 60 minute observation (not shown) has a low signal-to-noise ratio. The profile is broad and there appears to be a high fraction of linear polarisation although the RM cannot reliably be determined.

\noindent
{\bf J1755--0903} (Figure~\ref{fig1}): This pulsar's interpulse was not recognised at the time of discovery \citep{bbb+12}. The main pulse is triangular in shape. The circular polarisation changes sign in the centre of the profile and the linear polarisation profile is much narrower than in total intensity. The PA swing has a strange kink at the centre of the profile. The interpulse is broad, has much lower amplitude than the main pulse and its centre is only 165\degr\ from the main pulse centre.

\noindent
{\bf J1816--0755} (Figure~\ref{fig2}): Although discovered 15 years ago \citep{lsf+06} no polarisation data is available. The pulsar shows a narrow main and interpulse neither of which are highly polarised. Both main and interpulse appear to be blended doubles with the trailing component dominant. There is a very steep swing of PA across the main pulse.

\noindent
{\bf J1833--1034} (Figure~\ref{fig2}): This pulsar was found in a targetted search of the supernova remnant G21.5--0.9 \citep{crg+06} and has a low flux density. Two individual 60~min observations were combined. The low signal to noise ratio makes it hard to discern any features in the narrow profile. The polarisation fraction is high and the PA swing shallow.

\noindent
{\bf J1843--0702} (Figure~\ref{fig2}): The pulsar shows interpulse emission, with the peak separation of main and interpulse being 170\degr\ and the peak amplitudes about a factor 2 different. The profiles are narrow with a low degree of linear polarisation.

\noindent
{\bf J1849+0409} (Figure~\ref{fig2}): The main and interpulses in this pulsar have roughly similar amplitude. The main pulse has a triangular shape and is highly polarised with a simple PA swing. The interpulse is wider and has a double peaked profile with a hint of a central component. The linear polarisation is moderate and the PA swing rather flat.

\noindent
{\bf J1850--0026} (Figure~\ref{fig2}): The pulsar is highly scattered \citep{kel+09} but relatively bright. Both the linear and the circular polarisation fractions are high. The long, flat PA sweep is a by-product of the scattering process.

\noindent
{\bf J1856+0113} (Figure~\ref{fig2}): This pulsar has a previously published RM \citep{hml+06} but no published profile. Our observations show a narrow profile with two components. The linear polarisation fraction is moderate and the PA swing is flat. We have significantly reduced the error bar on the RM. In the compilation of \citet{wcl+99} there is a lack of linear polarisation in this pulsar, possibly due to an incorrect RM used at the time.

\noindent
{\bf J1856+0245} (Figure~\ref{fig2}): This pulsar is associated with a TeV source \citep{hng+08} but is not yet detected as a $\gamma$-ray pulsar. The pulse profile is broad and slightly scattered in the lower part of the frequency band. It has moderate linear and circular polarisation and there is a large gradient in PA across the profile.

\noindent
{\bf J1857+0143} (Figure~\ref{fig2}): This pulsar has a previously published RM \citep{hmvd18} but no published profile. Our RM (29.4$\pm$0.3 rad~m$^{-2}$) is not in agreement with the previous value (41.9$\pm$4.3 rad~m$^{-2}$). The pulsar is highly scattered at the low-end of the MeerKAT frequency band. At the high end of the band, we estimate a $W_{50}$ of $<25$\degr. The linear polarisation fraction is high. The long, flat PA sweep is a by-product of the scattering process.

\noindent
{\bf J1907+0631} (Figure~\ref{fig2}): The profile is broad and featureless and is virtually 100\% linearly polarised with a small fraction of negative circular polarisation. There is a large PA swing across the profile.

\noindent
{\bf J1907+0918} (Figure~\ref{fig3}): The profile is very narrow and has a high degree of linear and circular polarisation. The PA swing is steep likely indicating a central cut through the beam.

\noindent
{\bf J1909+0749} (Figure~\ref{fig3}): This pulsar was identified as having an interpulse by \citet{nab+13}. The main and interpulses have roughly equal amplitude but the interpulse is virtually 100\% linearly polarised unlike the main pulse. The main and interpulses have opposite signs of circular polarisation and opposite gradients in the PA swing.

\noindent
{\bf J1918+1541} (Figure~\ref{fig3}): The profile shows evidence for an interpulse, with low-amplitude emission located some 170\degr\ from the main pulse. The main pulse consists of two components, the leading component has low amplitude and almost no polarisation whereas the trailing component is bright with a high degree of both linear and circular polarisation.

\noindent
{\bf J1925+1720} (Figure~\ref{fig3}): The pulsar has a low flux density; its profile has a single component which is highly linearly polarised. 

\noindent
{\bf J1928+1746} (Figure~\ref{fig3}): The pulsar has a peculiar profile, with a small leading component separated from the main component which has a double structure. Linear polarisation is high, as is the circular polarisation under the main component. The PA swing indicates that there may be an orthogonal mode jump between the leading and main components.

\noindent
{\bf J1930+1852:} This pulsar is located in the supernova remnant SNR~54.1+0.3 and has a flux density of only 60~$\mu$Jy \citep{clb+02}. In spite of our deep integration, the signal to noise ratio is low in this profile and we have been unable to determine an RM. The profile (not shown) appears to be featureless and broad.

\noindent
{\bf J1932+2220} (Figure~\ref{fig3}): The pulsar has a narrow profile with a trailing shoulder. The polarisation fraction is high and there is a steep swing of PA. This is similar to the profile shown in \citet{wcl+99}. Our RM value (138.9$\pm$0.1 rad~m$^{-2}$) is not consistent with the previous value (173$\pm$11 rad~m$^{-2}$) listed in \citet{hl87}.

\noindent
{\bf J1934+2352} (Figure~\ref{fig3}): The profile of this pulsar consists of two blended components with the trailing component higher in amplitude. There is little circular polarisation and high linear polarisation with a shallow swing of PA.

\noindent
{\bf J1938+2213} (Figure~\ref{fig3}): The profile of this pulsar has 3 components with a shallow leading edge and a steep trailing edge. The linear polarisation is high against the first and third component but less so in the second component. There are two orthogonal mode jumps in the PA swing.

\noindent
{\bf J2043+2740} (Figure~\ref{fig3}): Discovered back in 1996 by \citet{rtj+96}, polarisation observations have only been carried out recently at low frequencies \citep{sbg+19}. Our observations show the profile has at least three components with the initial component dominating. The linear polarisation is very high and there is a steep swing of PA with an inflexion point late compared to the profile midpoint.

\section{Discussion}
\subsection{Interpulses}
We apply the rotating vector model (RVM) in order to derive the pulsar geometry from the PA values across pulse phase ($\phi$). We use a modified form of the original model of \citep{rc69} as presented in \citet{jk19b}.
\begin{equation}
\label{paswing2}
{\rm PA} = {\rm PA}_{0} +
{\rm arctan} \left( \frac{{\rm sin}\alpha
\, {\rm sin}(\phi - \phi_0 - \Delta)}{{\rm sin}\zeta
\, {\rm cos}\alpha - {\rm cos}\zeta
\, {\rm sin}\alpha \, {\rm cos}(\phi - \phi_0 - \Delta)} \right)
\end{equation}
Here, $\phi_0$ is the pulse longitude at which PA=PA$_{0}$ and $\zeta=\alpha+\beta$. The $\Delta$ term is present to deal with cases in which the emission heights are different for the main pulse and the interpulse. Details of the model fitting can be found in \citet{jk19b}. The results are presented in Table~\ref{tab:rvm}.
\begin{table*}
\caption{Results of fitting the RVM model to the interpulse pulsars. $\alpha_{M}$, $\beta_{M}$
and $\phi_{M}$ refer to the main pulse, $\alpha_{I}$, $\beta_{I}$
and $\phi_{I}$ to the interpulse.
$\phi$ is the location of the inflexion point of the PA swing with respect
to the peak of the main pulse emission. All angles in degrees.}
\label{tab:rvm}
\begin{tabular}{lrrrrrrrr}
Jname & $\alpha_{M}$ & $\beta_{M}$ & $\phi_{M}$ & $\zeta$ & $\alpha_{I}$ & $\beta_{I}$ & $\phi_{I}$ & $\Delta$ \\
\hline & \vspace{-3mm} \\
J0514--4408 &  65(2)   &  35(3)   &  32(2)   & 101(2)   & 115(2)   & --14(3)  & 191(4)   & --21(3)   \\
J1755--0903 &  83(1)   &   5(2)   & --1(1)   &  87(2)   &  97(2)   & --10(2)  & 174(2)   &  --5(2)   \\
J1816--0755 &  88(2)   & --3(2)   & --3(1)   &  85(2)   &  92(2)   & --7(3)   & 175(2)   &  --3(1)   \\
J1843--0702 &  91(1)   &   6(1)   &   3(1)   &  97(1)   &  88(1)   &   9(1)   & 180(2)   &  --3(1)   \\
J1849+0409  &  91.7(2) & --8.0(2) & --2.1(2) &  83.7(1) &  88.3(2) & --4.6(2) & 181.5(2) &    3.5(1) \\
J1909+0749  & 112(6)   & --32(8)  & --4(4)   &  80(7)   &  68(6)   &  12(9)   & 185(6)   &  --9(5)   \\
J1918+1514  &  93(1)   & --12(1)  & --8(1)   &  81(1)   &  87(1)   & --6(1)   & 164(2)   &  --8(1)   \\
\end{tabular}
\end{table*}

Five of the pulsars show `standard' values for orthogonal rotators (e.g. \citealt{jk19b}) with relatively small values of $\beta$ which, in combination with narrow pulse widths, implies `standard' emission heights of $\sim$300~km. However, there are two notable exceptions, namely PSRs~J0514$-$4408 and J1909$+$0749. 

The first peculiar feature of the fit to PSR~J0514$-$4408 is the  small $\alpha_M$ value of $65\degr\pm2\degr$, compared to our expectation of a value closer to 90\degr\ as seen for the other pulsars in Table~\ref{tab:rvm}. At the same time, however, $\beta_M$ is also unusually large, $35\degr\pm3\degr$, which in combination with $\zeta=101\degr\pm2\degr$ means an "inner line-of-sight",  as confirmed by the sweep of the PA swing \citep[cf.][]{lk05}. The large $\beta_M$ value immediately indicates that the beam radius $\rho_M$ must be large, as $|\beta_M|\lesssim\rho_M$ for the pulsar to be beamed at Earth.
This implies that the emission height of the main pulse is large. Using the relationship 
\begin{equation}
\label{height}
\rho = \sqrt{\frac{9\,\, h_{\rm em}}{4\,\,R_{\rm LC}}} \quad\mbox{\rm rad}
\end{equation}
\citep{ran90} where $R_{\rm LC}$ is the light cylinder radius, and using the observed $\beta_M$  value, we can estimate the main pulse emission height to be about $h_{{\rm em},M}\gtrsim2500$ km, i.e.~a unusually large fraction of the light-cylinder radius of 16\%. We can compare this with an emission height derived from the shift of the PA centroid relative to the pulse peak using the argument that this shift is caused by the rotating of the pulsar reference frame relative to the observer, which yields
a relationship that connects the phase shift with the emission height,
\begin{equation}
\label{bcw}
\delta\phi({\rm PA}) = \frac{4\,\,h_{\rm em}}{R_{lc}}
\end{equation}
\citep{bcw91,dyks08}. Identifying $\delta\phi({\rm PA})$ with our determined $\phi_M$, we derive a second estimate for the emission height of $2130\pm130$ km, which is remarkable consistent with the simple geometrical estimate from above. Turning our attention now to the interpulse, we see a more modest $\beta_{I}$ value of $-14$ deg, allowing the emission height to be much smaller, i.e.~$h_{{\rm em}, I}\sim400$ km. The large negative value of $\Delta = -21\pm3$ deg indeed suggests that the emission height of the interpulse is significantly lower than that of the main pulse. Interestingly, though, the width of the interpulse is 60 deg, which appears to be quite large. Indeed, taking the interpulse geometry and the relationship 
\begin{equation}
\label{rho}
{\rm cos}\rho = {\rm cos}\alpha\,\, {\rm cos}\zeta\,\, +\,\, {\rm sin}\alpha\,\, {\rm sin}\zeta\,\, {\rm cos}(W/2)
\end{equation}
\citep{ggr84}, where $W$ is the pulse width, the implied interpulse beam radius exceeds 80\degr. This is very difficult to reconcile with the lower emission height just derived. We have to conclude that while we can find a consistent description of the geometry for the main pulse, albeit with an unusually large emission height, the large width of the interpulse is inconsistent with the lower emission height that is implied by the RVM fit. We will return to the implication of large radio pulse widths in the next section in the context of gamma-ray emission.

The second pulsar with a rather large impact value is PSR~J1909$+$0749, where we find $\beta_M=-32\degr\pm8\degr$, again implying an inner line of sight. Repeating previous arguments, we suppose that $\rho_M>|\beta_M|$, which implies an emission height exceeding 1500~km. That is much larger as one would infer from the low $\phi_M$ value, which would require a height of only $\sim220$~km. A negative value of $\Delta$ implies an even smaller emission height for the interpulse, but the relative uncertainty is large. The pulse widths for main and interpulse are similar, about 17\degr. For the interpulse geometry, we can obtain a beam radius of 15\degr\,consistent with a value of $\sim$13\degr\,expected from a simple period scaling. In contrast, for the main pulse geometry, we derive a beam radius that is about twice larger because of the large $\beta_M$ value. In summary, for this pulsar, we can obtain a consistent picture for the interpulse, while for the main pulse the large impact angle of the main pulse derived from the RVM fit is more difficult to reconcile with the other observed pulse properties.

A correlation between the sign of the circular polarisation in the main and interpulse and the sign of $\beta$ was noted by \citet{jk19b}. For the cases under consideration here, only PSR~J1909+0749 gives clear indication of circular polarisation under both poles. This pulsar obeys the correlation; $\beta$ has opposite sign in the main and interpulse and so does the circular polarization. PSR~J1755--0903 is a rare example where the sign of circular polarization changes under the main pulse.

\subsection{Radio profile widths, geometry and $\gamma$-ray emission}
\begin{figure}
\begin{center}
\includegraphics[width=8cm,angle=0]{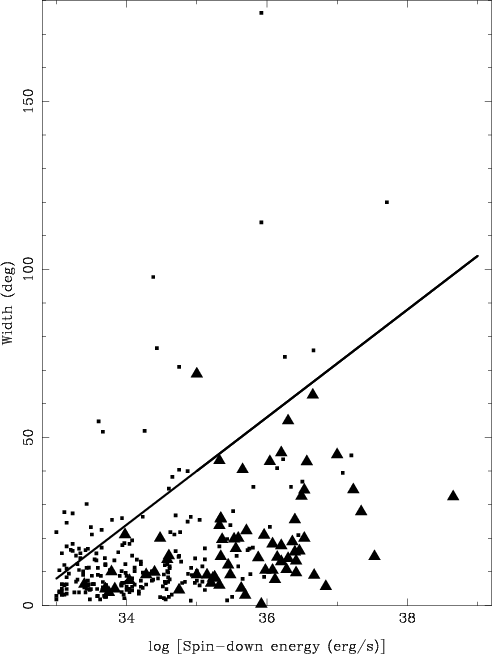}
\end{center}
\caption{Spin-down energy versus radio pulse width for a sample of radio-only pulsars (squares) and joint $\gamma$-ray and radio pulsars (triangles). The line represents demarcation as defined by equation~\ref{eq:rook}, above which no $\gamma$-ray pulsars are expected. The one exception is PSR~J0631+0646.}
\label{fig:edot-width}
\end{figure}
\citet{rwjk17} showed that it is possible to distinguish between pulsars which are seen in radio but not $\gamma$-rays and those which are visible at both wavebands based on the width of the radio profile ($W$). In addition, the modelling in \citet{jskk20} showed that $\gamma$-ray pulsars are not expected above a line which follows:
\begin{equation}
    W = 16.0\,\,\, {\rm log} \left(\frac{\dot{E}}{10^{35}{\rm erg\;s^{-1}}}\right) + 40.0
\label{eq:rook}
\end{equation}
with $W$ in degrees. In addition, the outer-gap model predicts that at lower values of $\dot{E}$, $\gamma$-ray pulsars must be close to orthogonal rotators \citep{wrwj09}.  We show an update of the \citet{rwjk17} results in Figure~\ref{fig:edot-width} along with the demarcation line given by equation~\ref{eq:rook}. Values of $W_{50}$ are taken from \citet{jk18}, the results presented in this paper and other measurements from the literature.

The following aspects can be noted from the figure. First, there is only one $\gamma$-ray pulsar above the line of equation~\ref{eq:rook}, PSR~J0631+0646. As described above, the pulsar has a profile with two widely separated components and a flat swing of PA likely indicating a low value of $\alpha$. It is therefore unclear why this pulsar should be $\gamma$-ray bright. For $\dot{E}<10^{35}$~erg~s$^{-1}$, a sizeable fraction of the radio population lies above the line, but no $\gamma$-ray pulsars do. Radio pulsars at high $\dot{E}$ with very wide profiles (e.g. PSR~J1302--6350) are not seen in $\gamma$-rays. Finally for $\dot{E}>10^{35}$~erg~s$^{-1}$ the two populations of pulsars seem more intermixed than was the case in \citet{rwjk17}, though we note that \citet{rwjk17} used $W_{10}$ rather than $W_{50}$.

\subsection{$\gamma$-ray pulsar candidates}
\begin{table}
\caption{Estimated $\gamma$-ray fluxes of the sample of radio-only pulsars.}
\label{tab:fg}
\begin{tabular}{lc || lc}
Jname & $F_g \times 10^{-12}$ &
Jname & $F_g \times 10^{-12}$ \\
      & (ergs$^{-1}$cm$^{-2}$) &
      & (ergs$^{-1}$cm$^{-2}$) \\
\hline & \vspace{-3mm} \\
J1755--0903     &   410 &
J1437--5959     &     5 \\
J0835--3707     &    53 &
J1849+0409      &     5 \\
J1400--6325     &    47 &
J1850--0026     &     4 \\
J1747--2809     &    33 &
J1909+0749      &     3 \\
J1930+1852      &    23 &
J1906+0746      &     3 \\
J1918+1541      &    23 &
J1907+0918      &     3 \\
J1907+0631      &    21 &
J1934+2352      &     2 \\
J1856+0245      &    18 &
J1843--0702     &     2 \\
J1938+2213      &    17\\
\end{tabular}
\end{table}

For a radio pulsar to be seen in $\gamma$-rays it is necessary that the $\gamma$-ray flux, $F_g$, exceed the Fermi-LAT threshold over the lifetime of the mission. In the outer-gap model $F_g$ is given by
\begin{equation}
F_g = \frac{1}{4\pi c_g} \,\, \frac{\dot{E}}{d^2} \sqrt{\frac{10^{33}}{\dot{E}}}
\label{eqn:fg}
\end{equation}
\citep{wrwj09}, where $(4\pi c_g)^{-1}$ is a geometric term, which is $\sim$0.1 for $\alpha>60$\degr\ \citep{jskk20} and $d$ is the distance to the pulsar. Using $\dot{E}$ and $d$ from Table~\ref{tab_results}, we compute $F_g$ for each of the 17 radio-only pulsars in our sample and list the results in Table~\ref{tab:fg}. 

The detection threshold of the Fermi-LAT depends on the Galactic location of the pulsar and whether or not radio timing can provide a coherent timing ephemeris since the start of the Fermi mission in 2008 \citep{sgc+08,sbc+19}. Unfortunately none of the pulsars listed here have sufficient timing accuracy to allow this, thus raising the detection level. We therefore take the sensitivity to pulsars at low Galactic latitudes to be $16\times10^{-12}$~erg cm$^{-2}$s$^{-1}$ for pulsars without a coherent ephemeris, scaled from \citet{2pc} to allow for increased time span (see also \citealt{jskk20}). Pulsars on the left of Table~\ref{tab:fg} are therefore above the detection threshold whereas those on the right are below the detection limit.

Of the 8 pulsars below the detection limit, three are interpulses, PSRs~J1843--0702, J1849+0409 and J1909+0749. Although the former two pulsars are relatively nearby, their low $\dot{E}$ counts against them, whereas the high $\dot{E}$ of PSR~J1909+0749 is mitigated by its large distance.

Of the 9 pulsars nominally above the detection limit, three are interpulses, PSRs~J0835--3707, J1755--0903 and J1918+1541. None appear in the $\gamma$-ray source catalogue (4FGL; \citealt{4FGL}). This is very surprising for PSRs~J0835--3707 and J1755--0903 which are both nearby and in relatively quiet parts of the Galactic plane. They have $\dot{E}$ similar to, but distances less than PSR~J0514--4408 which is detected in $\gamma$-rays.

PSRs~J1907+0631 and J1938+2213 are similar in that they both have moderate width profiles and a steep swing of PA. The geometry is unclear. Both are therefore potential $\gamma$-ray candidates. The catalogued source 4FGL~J1906.2+063, tentatively associated with SNR~G40.5--0.5, lies close to PSR~J1907+0631 on the sky which may make the pulsar harder to detect.

PSRs~J1930+1852 and J1856+0245 have wide profiles, likely indicating a small value of $\alpha$. Their widths lie above the line in Figure~\ref{fig:edot-width} and they are therefore unlikely to be seen as $\gamma$-ray pulsars. We note that a fit to the X-ray torus around PSR~J1930+1852 \citep{ngr08} yields $\alpha+\beta=147$\degr meaning that for small $\beta$ the pulsar could indeed be far from orthogonality.

Similarly, PSR~J1400--6325 has an extremely broad radio profile and its $W_{50}$ places the pulsar well above the line in Figure~\ref{fig:edot-width}. This would make it unlikely to be seen in $\gamma$-rays. However, X-ray pulsations are detected from this pulsar. This source appears to be very similar to PSR~J1513--5908 which is also a bright X-ray pulsar with a broad radio profile albeit detected in the $\gamma$-ray. Determining the geometry of PSR~J1400--6325 would be informative but we are unable to do so with the current radio data. It is possible that the geometry of PSR~J1400--6325 could be discerned from X-ray data; it lies within a pulsar wind nebula although the presence of an X-ray torus is not clear \citep{rb19}.

In summary, of the 17 radio-only pulsars in this sample, 8 are likely below the sensitivity of the Fermi-LAT and a further 3 have non-favourable geometries. The failure to detect the interpulse objects PSRs~J0835--3707 and J1755--0903 in $\gamma$-rays is surprising with more detailed modelling needed to understand the reasons behind this.

\subsection{Linear polarisation fraction and $\gamma$-ray emission}
\citet{wj08b} and later \cite{jk18} showed that there was a correlation between the fractional linear polarisation of a pulsar's profile and $\dot{E}$. For $\dot{E}>10^{35}$~erg~s$^{-1}$, the fractional polarisation is well above 40\% in the vast majority of the pulsars, whereas for $\dot{E}<10^{34}$~erg~s$^{-1}$ the fractional polarisation is lower. There is a transition in the fractional polarisation levels between $10^{34} < \dot{E} < 10^{35}$~erg~s$^{-1}$. One possible explanation for this transition is that the high $\dot{E}$ pulsars have high emission heights in the radio, allowing the polarisation to escape the magnetosphere whereas the emission from low $\dot{E}$ pulsars originates much deeper in the magnetosphere where propagation effects become important \citep{kj07}.

It is well known that there is also a strong correlation between $\dot{E}$ and $\gamma$-ray luminosity \citep{2pc}. Only 17 young pulsars with $\dot{E}<10^{35}$~erg~s$^{-1}$ are known to emit both radio and $\gamma$-rays in spite of extensive $\gamma$-ray searches \citep{sbc+19}. The $\gamma$-ray emission certainly occurs high in the magnetosphere (e.g. \citealt{rom96,ps18}). Is it possible then that there is some causal link between high linear polarisation in the radio and the visibility of a pulsar in $\gamma$-rays?

Of the 37 pulsars with $\dot{E}>10^{35}$~erg~s$^{-1}$ detected only in radio and with polarisation measured, only 5 have a fractional polarisation less than 40\%. There are 64 pulsars above this $\dot{E}$ with both radio and $\gamma$ emission. Of the 56 with measured polarisation, 5 have a low fraction. In these two samples therefore, 86\% of the radio-only population have high linear polarisation as opposed to 91\% for the $\gamma$-ray plus radio pulsars.
Finally, there are 17 pulsars with $\dot{E}<10^{35}$~erg~s$^{-1}$ which are both radio and $\gamma$-ray emitters of which 14 now have their polarisation properties measured. Of these 14, 8 have high polarisation and 6 have low polarisation. This is consistent with the ratio of high to low polarisation in the radio-only population below this $\dot{E}$.

Taken together, these numbers indicate that the joint radio and $\gamma$-ray pulsar population show the same polarisation properties as the radio-only population. Furthermore, if high polarisation in the radio implies a high emission height then this appears not to be a pre-requisite to engage the $\gamma$-ray engine. This makes the link between high polarization in the radio and $\gamma$-ray unclear. However, a causal correlation between the two remains an option; it is possible that the high altitude cascades that yield $\gamma$-rays occur when there is also high altitude radio emission.

\section{Summary}
We have exploited the excellent sensitivity of the MeerKAT telescope to observe a sample of high $\dot{E}$ pulsars with low flux densities and without polarisation profiles in the literature. We show a correlation between the widths of the radio profiles and $\gamma$-ray detectability. We use this correlation to surmise that pulsars from our sample could be seen in $\gamma$-rays if a sustained timing campaign were to be carried out. Although high $\dot{E}$ radio pulsars are highly polarised and high $\dot{E}$ pulsars are more likely to be seen in $\gamma$-rays, there appears to be no obvious link between these two observables. Not all $\gamma$-ray pulsars have highly polarised radio emission. The geometry of a pulsar is important, especially at low $\dot{E}$ where 4 out of 17 pulsars which show both radio and $\gamma$-ray emission are orthogonal rotators.

\section*{Acknowledgements}
We thank the referee Dr David Smith for extremely constructive suggestions that resulted in the improvement of the paper. The MeerKAT telescope is operated by the South African Radio Astronomy Observatory, which is a facility of the National Research Foundation, an agency of the Department of Science and Innovation. MeerTime data is housed and processed on the OzSTAR supercomputer at Swinburne University of Technology with the support of ADACS and the gravitational wave data centre via AAL. This research was funded partially by the Australian Government through the Australian Research Council, grants CE170100004 (OzGrav) and FL150100148. RMS acknowledges support through Australian Research Council Future Fellowship FT190100155.

\section*{Data Availability}
The data underlying this article will be shared on reasonable request to the corresponding author.

\bibliographystyle{mnras}
\bibliography{mk_gray} 

\bsp	
\label{lastpage}
\end{document}